\documentclass[aps,prl,twocolumn,twoside,superscriptaddress,nofootinbib,preprintnumbers]{revtex4-1}

\usepackage{hyperref}
\usepackage{graphicx}
\usepackage{amsmath}
\usepackage{mathrsfs}
\usepackage{xspace}
\usepackage{color}
\usepackage[normalem]{ulem}

\definecolor{darkgreen}{rgb}{0.13,0.55,0.13}

\newcommand{\eq}[1]{Eq.~\eqref{eq:#1}}

\newcommand{\fig}[1]{Fig.~\ref{fig:#1}}

\newcommand{\df}{\mathrm{d}}

\newcommand{\nn}{\nonumber}

\allowdisplaybreaks[2]

\begin{document}


\title{Collinear Parton Dynamics Beyond DGLAP}

\author{Hao Chen}
\affiliation{Zhejiang Institute of Modern Physics, Department of Physics, Zhejiang University, Hangzhou, 310027, China}

\author{Max Jaarsma}
\affiliation{Nikhef, Theory Group,
	Science Park 105, 1098 XG, Amsterdam, The Netherlands}
\affiliation{Institute for Theoretical Physics Amsterdam and Delta Institute for Theoretical Physics, University of Amsterdam, Science Park 904, 1098 XH Amsterdam, The Netherlands}

\author{Yibei Li}
\affiliation{Zhejiang Institute of Modern Physics, Department of Physics, Zhejiang University, Hangzhou, 310027, China}

\author{Ian Moult}
\affiliation{Department of Physics, Yale University, New Haven, CT 06511}

\author{Wouter J.~Waalewijn}
\affiliation{Nikhef, Theory Group,
	Science Park 105, 1098 XG, Amsterdam, The Netherlands}
\affiliation{Institute for Theoretical Physics Amsterdam and Delta Institute for Theoretical Physics, University of Amsterdam, Science Park 904, 1098 XH Amsterdam, The Netherlands}

\author{Hua Xing Zhu}
\affiliation{Zhejiang Institute of Modern Physics, Department of Physics, Zhejiang University, Hangzhou, 310027, China}

\begin{abstract}
Renormalization group evolution equations describing the scale dependence of quantities in quantum chromodynamics (QCD)  play a central role in the interpretation of experimental data. 
Arguably the most important evolution equations for collider physics applications are the Dokshitzer-Gribov-Lipatov-Altarelli-Parisi (DGLAP) equations,  which describe the evolution of a  quark or gluon fragmenting into hadrons, with only a \emph{single} hadron identified at a time.
In recent years, the study of the correlations of energy flow within jets has come to play a central role at collider experiments,  necessitating an understanding of correlations, going beyond the standard DGLAP paradigm.
In this \emph{Letter} we derive a general renormalization group equation describing the collinear dynamics that account for correlations in the fragmentation.
We compute the kernel of this evolution equation at next-to-leading order (NLO), where it involves the $1\to 3$ splitting functions, and develop techniques to solve it numerically. 
We show that our equation encompasses all previously-known collinear evolution equations, namely DGLAP and the evolution of multi-hadron fragmentation functions.
As an application of our results, we consider the phenomenologically-relevant example of energy flow on charged particles, computing the energy fraction in charged particles in $e^+e^- \to$ hadrons at NNLO.
Our results are an important step towards improving the understanding of the collinear dynamics of jets, with broad applications in jet substructure, ranging from the study of multi-hadron correlations, to the description of inclusive (sub)jet production, and the advancement of modern parton showers.

\end{abstract}

\maketitle

\emph{Introduction.}---Jets and their substructure play a central role in modern collider experiments, both in searches for beyond the Standard Model physics, as well as for studying quantum chromodynamics (QCD) \cite{Larkoski:2017jix,Asquith:2018igt,Marzani:2019hun}. Due to the confinement process, jets are complicated multi-scale objects, formed by the fragmentation of an initial energetic quark or gluon at short times, into a collimated spray of hadrons at long times. Because of this multi-scale nature, renormalization group equations (RGE) that describe the scale evolution of jets play a crucial role in the interpretation of nearly all experimental data.  

The most celebrated evolution equation is the Dokshitzer-Gribov-Lipatov-Altarelli-Parisi (DGLAP) \cite{Gribov:1972ri,Dokshitzer:1977sg,Altarelli:1977zs} equation, describing the evolution of a quark or gluon fragmenting into hadrons, with a single hadron identified at a time and all hadrons summed over. This equation plays a central role in the description of all aspects of jets, from perturbative calculations, to parton shower algorithms, to the evolution of fragmentation functions. Because of this, it has received significant theoretical attention, and been computed to high orders \cite{Mitov:2006ic,Mitov:2006wy,Chen:2020uvt}.

\begin{figure}
  \centering
    \includegraphics[width=0.45\textwidth]{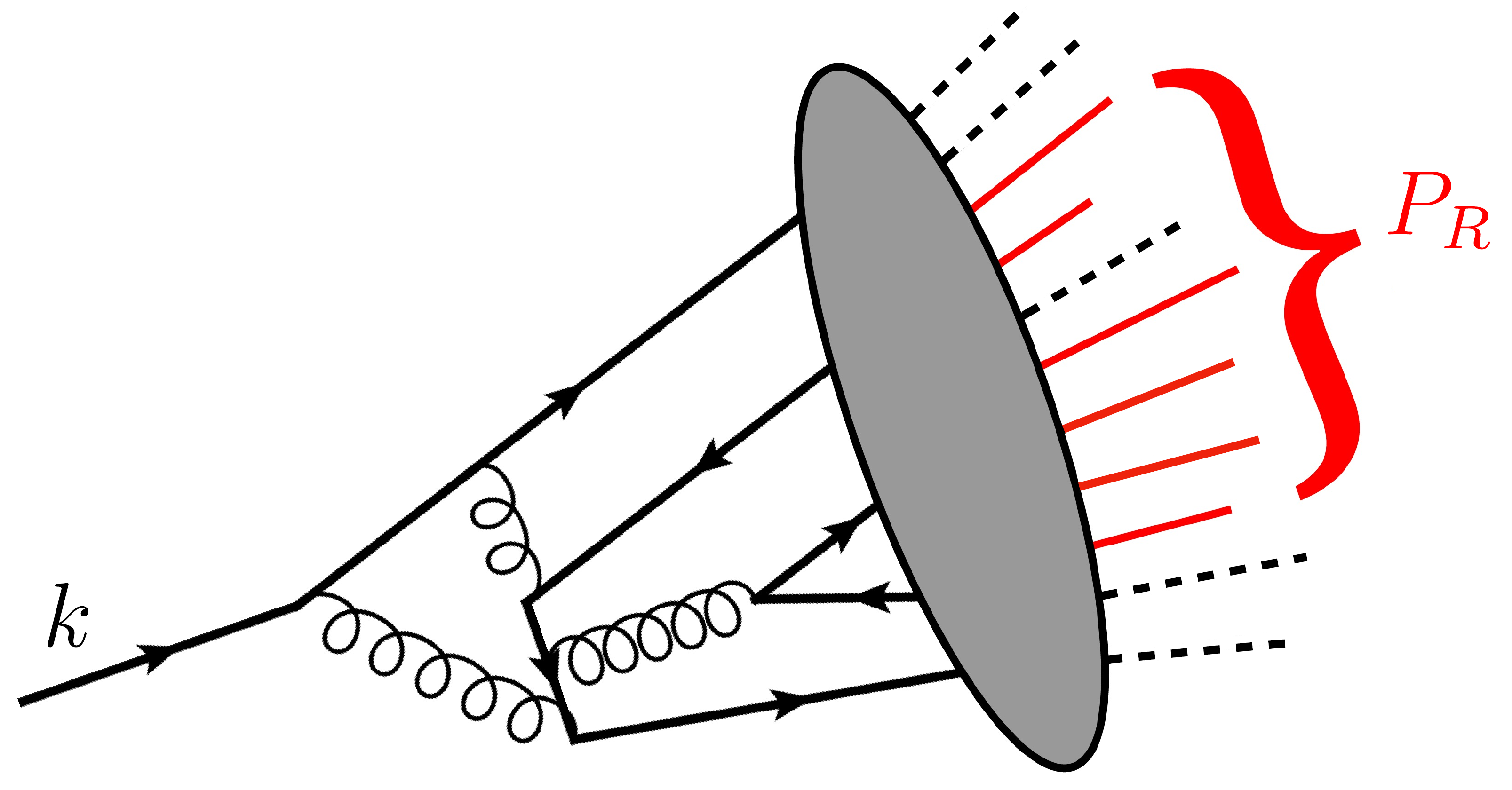}
  \caption{A parton with momentum $k$ fragments into an identified set of hadrons with momentum fraction $P_R$, distinguished by a specified quantum number (e.g.~electric charge). The scale evolution of this process is described by a non-linear renormalization group evolution.}
  \label{fig:schematic}
\end{figure}

Driven by the high energies, and exceptional resolution of the detectors at the Large Hadron Collider (LHC), there has been significant recent interest in understanding the correlations in energy flow \emph{within} jets, a field known as jet substructure \cite{Larkoski:2017jix,Asquith:2018igt,Marzani:2019hun}. The theoretical description of such correlations requires an understanding of the scale evolution of correlations in the fragmentation process, giving rise to non-linear RGEs, and going beyond the standard DGLAP evolution equations. While non-linear evolution equations for soft correlations have existed for quite some time \cite{Dasgupta:2001sh,Banfi:2002hw,Hatta:2013iba,Caron-Huot:2015bja,Banfi:2021owj,Banfi:2021xzn},  similar non-linear evolution equations incorporating correlations between collinear partons are not known. Much like their soft analogs, such non-linear collinear evolution equations will also be essential for testing higher order collinear corrections to next-generation parton showers \cite{Li:2016yez,Hoche:2017hno,Hoche:2017iem,Dulat:2018vuy,Gellersen:2021eci,Hamilton:2020rcu,Dasgupta:2020fwr,Hamilton:2021dyz,Karlberg:2021kwr}.

In this \emph{Letter} we  derive a general evolution equation for the fragmentation of collinear partons at next-to-leading order (NLO), accounting for all correlations, which involves for the first time the complete structure of the $1\to 3$ splitting functions. We also show that our evolution equation can be reduced to the DGLAP equation, as well as the $N$-hadron fragmentation functions. 

There are many applications of our evolution equations for jet substructure at the LHC. A key application is the study of observables on tracks (charged particles)~\cite{Chang:2013rca,Chang:2013iba}, which will enable the measurement of more sophisticated jet substructure observables.  In particular, the use of tracks has played an important role in the study of the collinear limit of energy correlators~\cite{Dixon:2019uzg,Chen:2019bpb,Chen:2020vvp} using CMS Open Data~\cite{Komiske:2022enw,Chen:2022swd}. This same equation can be used to describe (sub)jet production for small values of the radius, e.g.~to describe the leading (sub)jet requires knowledge about the other (sub)jets~\cite{Scott:2019wlk,Neill:2021std}.
We therefore develop algorithms to efficiently solve our equations numerically for phenomenological applications, and illustrate their use in a physical observable, the energy fraction in charged hadrons in an $e^+e^-$ collision.

\begin{figure}[t]
  \centering
    \includegraphics[width=0.44\textwidth]{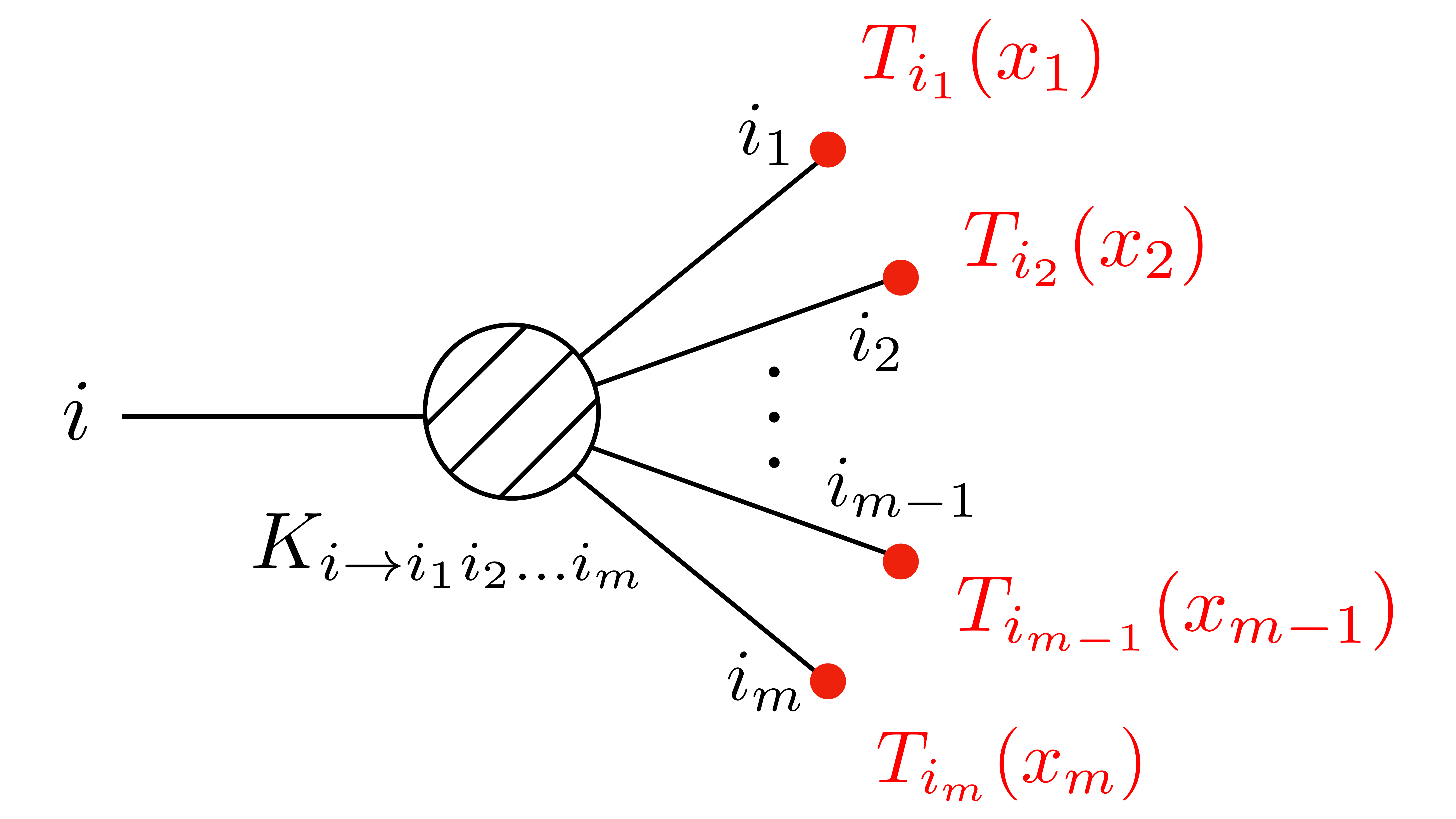}
  \caption{Perturbative evolution is described by the splitting of a single parton, $i$, into $m$ partons, each of whose momentum fractions are tracked.}
  \label{fig:nhadron}
\end{figure}

\emph{The Master Equation for Collinear Evolution.}---We derive the general collinear evolution equation by studying the renormalization of a universal object referred to as a ``track function" \cite{Chang:2013rca,Chang:2013iba}. In light-cone gauge, it is defined for quarks  as \cite{Chang:2013rca,Chang:2013iba}
\begin{align} \label{T_def}
T_q(x)&=\!\int\! \df y^+ \df ^{d-2} y_\perp e^{ik^- y^+/2} \sum_X \delta \biggl( x\!-\!\frac{P_R^-}{k^-}\biggr)\nn \\
&\quad \frac{1}{2N_c}
\text{tr} \biggl[  \frac{\gamma^-}{2} \langle 0| \psi(y^+,0, y_\perp)|X \rangle \langle X|\bar \psi(0) | 0 \rangle \biggr]\,,
\end{align} 
with a similar definition for gluons.
Here $P_R$ denotes the momentum of a subset $R \subset X$ of hadrons, as illustrated in \fig{schematic}. The RGE of these universal non-perturbative quantities tracks the energy fractions of all partons in a splitting, as is shown schematically in \fig{nhadron}, leading to a complicated non-linear evolution equation, whereas DGLAP considers the momentum fraction of one parton produced in a splitting at the time, summing over them. In \cite{Li:2021zcf,Jaarsma:2022kdd}, it was shown how to derive RG equations for the first six moments of the track functions. In this \emph{Letter}, we will extend this to derive the complete RGE at NLO.

Considering the perturbative splitting illustrated in \fig{nhadron}, at NLO, we can have a splitting into at most three-partons. The general form of the evolution equation for the track functions is therefore 
\begin{align}
 \frac{\df }{\df \ln\mu^2}T_i(x) &= a_s \Bigl[
 K^{(0)}_{i\to i} T_i(x)  +
 K^{(0)}_{i\to i_1i_2}\otimes T_{i_1}T_{i_2}(x) \Bigr]
 \nn \\ & \quad
 +
a_s^2\Bigl[
K^{(1)}_{i\to i} T_i(x) 
+
K^{(1)}_{i\to i_1i_2}\otimes T_{i_1}T_{i_2}(x)
 \nn \\
  & \quad  +K^{(1)}_{i\to i_1i_2i_3}\otimes T_{i_1} T_{i_2} T_{i_3}(x)
  \Bigr] + \mathcal{O}(a_s^3)\,,
\end{align}
where $a_s = \alpha_s/(4\pi)$ is the coupling and the convolutions involving the evolution kernels $K$ are written explicitly below in \eq{N_is_4}.

Since track functions are scaleless in dimensional regularization, to derive their evolution equations, we follow the approach of \cite{Li:2021zcf,Jaarsma:2022kdd} of computing an auxiliary observable which introduces a scale. We take this to be a jet function differential in both the energy fraction of charged particles, and the mass of all particles~\cite{Ritzmann:2014mka}. After renormalization in the mass, the remaining poles are associated with the track function renormalization. 

To derive the evolution equations, one must integrate out the angular variables appearing in the $1\to 3$ splitting function \cite{Kosower:2003np}. The derivation of IR finite evolution equations then requires the cancellation poles between the one-loop $1\to 2$ \cite{Campbell:1997hg,Catani:1998nv} and tree-level $1\to 3$ \cite{Campbell:1997hg,Catani:1998nv} splitting functions. This is non-trivial as they can be overlapping so that standard plus distributions cannot be used. To overcome this, we use sector decomposition \cite{Binoth:2000ps,Binoth:2004jv,Anastasiou:2003gr}, and identify a convenient set of variables which disentangle the divergences.

We have computed the evolution kernels in both  $\mathcal{N}=4$ super-Yang-Mills (SYM) and for all partonic channels in QCD. The QCD kernels are provided in an attached notebook. We will use the simpler $\mathcal{N}=4$ SYM kernels to illustrate the form of the track function evolution,
\begin{widetext}
\begin{align} \label{eq:N_is_4}
  &\frac{\df}{\df \ln\mu^2}T(x)= 
 -25\zeta_3\, a^2 \, {\color{red}T(x)} + 
  \int_0^1\df x_1\int_0^1\df x_2\int_0^1\df z\ {\color{red}T(x_1)T(x_2)}
  \ \delta\left(x-x_1\frac{1}{1+z}-x_2\frac{z}{1+z}\right) \\
  &\hspace{2.5cm}\times \biggl\{4a \biggl[\frac{1}{z}\biggr]_+ + 16 a^2 \biggl( \zeta_2
  \biggl[\frac{1}{z}\biggr]_+  + \frac{2 \ln^2(1\!+\!z)-\ln z \ln(1\!+\!z)}{z} \biggr)  \biggr\}\nn\\
  &+ 8a^2\int_0^1\!\df x_1 \int_0^1\! \df x_2 \int_0^1\! \df x_3 \int_0^1\! \df z \int_0^1\!\df t\ {\color{red} T(x_1)T(x_2)T(x_3)}
  \ \delta\Bigl(x-x_1\frac{1}{1\!+\!z\!+\!zt}-x_2\frac{z}{1\!+\!z\!+\!zt}-x_3\frac{zt}{1\!+\!z\!+\!zt}\Bigr)\nn\\
  &\times\!
   \biggl\{\frac{4  \ln
   (1\!+\!z)}{z}\biggl[\frac{1}{t}\biggr]_+ \hspace{-0.3cm}+\biggl[\frac{1}{z}\biggr]_+\hspace{-0.2cm} \left(4
   \left[\frac{\ln t}{t}\right]_+\hspace{-0.3cm} -\frac{\ln t}{1\!+\!t}-\frac{7 \ln
   (1\!+\!t)}{t}\right) 
   +\frac{2 \left[ \ln (1\!+\!t z)-\ln (1\!+\!z\!+\!t z)\right]}{(1\!+\!t)(1\!+\!z) (1\!+\! t z)}
   +\frac{10 \left[\ln (1\!+\!z\!+\!t z)-\ln (1\!+\!z) \right]}{t z} 
   \nn\\
   & \quad  -\frac{7 \ln (1\!+\!t z)}{tz}+\frac{\ln (1\!+\!t)-\ln t}{(1\!+\!t) (1\!+\!t z)}+\frac{\ln(1\!+\!z)+\ln (1\!+\!t)}{(1\!+\!t)(1\!+\!z)}
   -\frac{\ln (1\!+\!z)}{(1\!+\!t) z}
   -\frac{z \ln (1\!+\!z)}{(1\!+\!z) (1\!+\!t z)}
   +\frac{\ln (1\!+\!t z)}{(1\!+\!t) z (1\!+\!z)}
  \biggr\}+\mathcal{O}(a^3)\,.\nn
\end{align}
\end{widetext}
As expected, this is a non-trivial kernel describing the mixing of a single track function into products of up to three track functions at $\mathcal{O}(a^2)$. It takes a relatively simple form involving plus distributions and logarithms.
Interestingly, it exhibits a form of maximal transcendentality, which would be interesting to understand better.

\begin{figure}[b]
  \centering
 \includegraphics[width=0.45\textwidth]{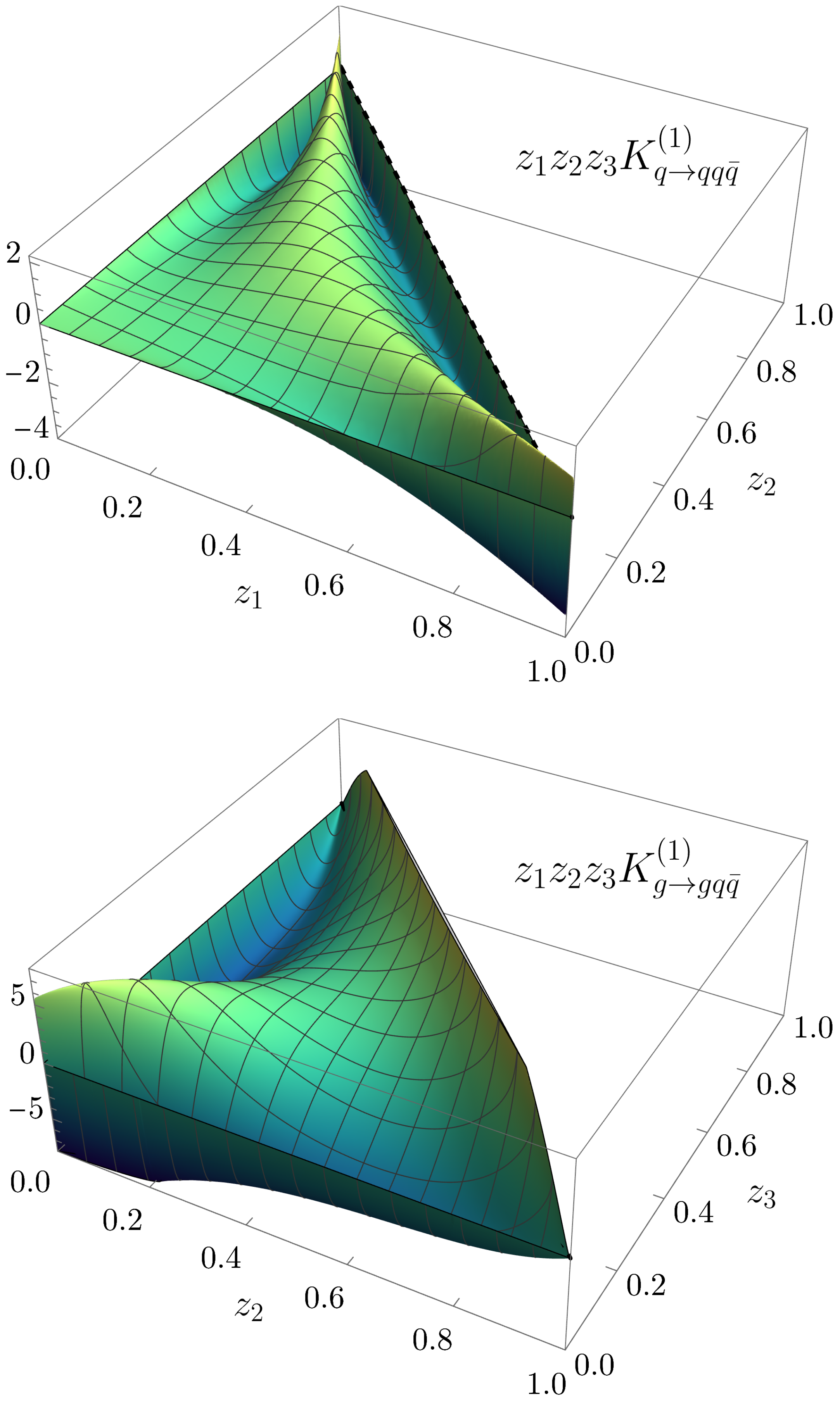} 
  \caption{The evolution kernels $K^{(1)}_{q\to qq\bar q}$ and $K^{(1)}_{g\to g q\bar q}$ as a function of the energy fractions. We have multiplied by $z_1 z_2 z_3$ to remove singularities at the phase space boundaries. The kernels exhibit non-trivial kinematic features reflecting how energy is distributed amongst correlated fragmenting partons.}
  \label{fig:ggg}
\end{figure}

Since the expressions for the QCD evolution kernels are more lengthy, we illustrate our results by plotting the evolution kernels for $g\to g q \bar q$ and $q \to q q \bar q$ in \fig{ggg} as a function of the momentum fractions (Note $z_1+z_2+z_3=1$). These kernels exhibit interesting features describing correlations in the energy distribution of fragmenting partons.

To our knowledge, these are the first complete set of $1\to 3$ collinear evolution equations, and we believe that they will play an important role in future studies of jet substructure. It is interesting to note that there exist early attempts at $1\to 3$ evolution equations \cite{Gunion:1984xw,Gunion:1985pg} in the framework of the jet calculus \cite{Konishi:1978yx,Konishi:1979cb}. However, we were unable to relate those approaches to our results.

We have performed a number of highly non-trivial checks on our results. First, we have checked that the first six moments of the QCD result agree with our results computed previously \cite{Li:2021zcf,Jaarsma:2022kdd}. Secondly, we have checked it against the known DGLAP results by integrating out momentum fractions, as will be described in the next section.

\emph{Reduction to DGLAP and $N$-Hadron Fragmentation.}---Since our evolution equation tracks the full momentum dependence of the partons, it can be viewed as the most general evolution equation for collinear dynamics. It is therefore interesting to show how it encodes the standard DGLAP evolution equation, as well as the evolution of the $N$-hadron fragmentation functions \cite{Konishi:1979cb,Sukhatme:1980vs,Sukhatme:1981ym,Vendramin:1981te,deFlorian:2003cg,Majumder:2004wh} for all values of $N$. Our results thus also yield the evolution equations for $N$-hadron fragmentation functions at NLO, which were not known yet. These objects are themselves of interest in jet substructure, or for studying modifications of the hadronization process in heavy ion collisions \cite{Majumder:2005ii}\cite{Majumder:2008jy}.

A key difference between our evolution equation and that of $N$-hadron fragmentation, is that by tracking the energy fraction of each final state particle in the splitting, energy conservation is manifest, while the standard DGLAP evolution arises from radiative energy loss, since one tracks only a finite number of the final state hadrons. However, this suggests a simple way of deriving the the kernels for $N$-hadron fragmentation functions from our kernels, namely by ``integrating out" some subset of the particles. 

\begin{figure}[t]
  \centering
    \includegraphics[width=0.49\textwidth]{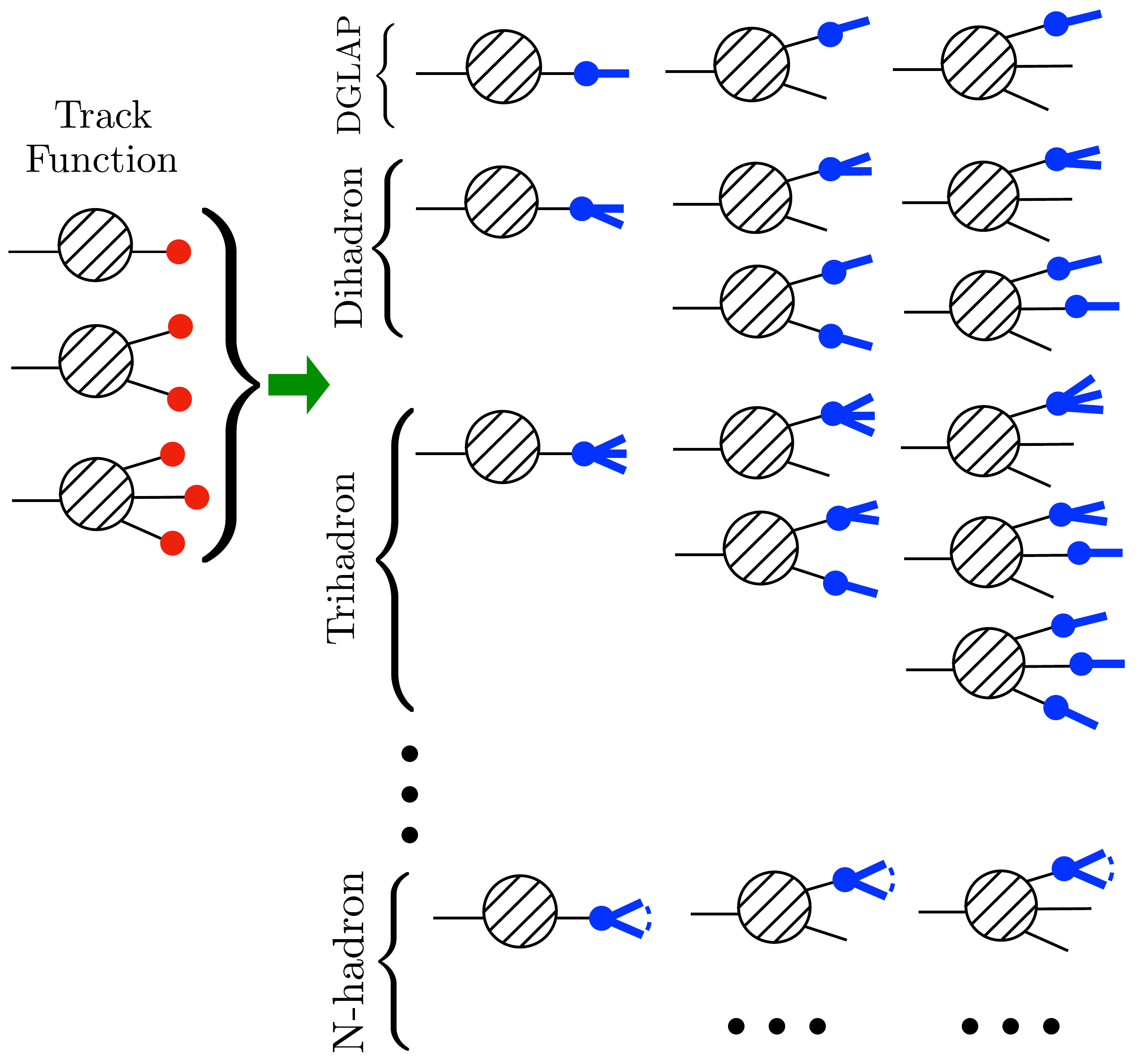}
  \caption{The single evolution equation for the track function can be repackaged as an infinite set of evolution equations for the $N$-hadron fragmentation functions, by integrating out energy fractions. This is illustrated for single hadron, di-hadron, and tri-hadron, with a straightforward extension to $N$-hadron. Permutations of diagrams are not shown.}
  \label{fig:reduction}
\end{figure}

The more precise relation is shown in \fig{reduction}. To reduce to DGLAP (single hadron fragmentation), one must integrate out all particles except one.  This can be done by substituting $T(x_i)\to \delta(x_i)$ for the corresponding lines. Note that only one representative diagram is shown and that the particle that is singled out can correspond to any of the lines, and one must sum over all possibilities.
This procedure extends to higher $N$-hadron fragmentation functions, as illustrated for the di-hadron case in \fig{reduction}, where one integrates out all particles except subsets of $1$ or $2$. This illustrates that our equation can be viewed as a repackaging of the infinite set of $N$-hadron fragmentation functions into a single equation.

\emph{Numerical Solutions of the Evolution Equations.}---The collinear renormalization group equations derived above are non-trivial coupled integro-differential equations. We will therefore show that they can be solved numerically, extending the approaches used for the leading order evolution in Refs.~\cite{Chang:2013rca,Chang:2013iba}.
We have developed several complementary algorithms to solve the evolution equations, to ensure consistency. Due to the presence of distributions in the evolution equations, it is convenient to perform an integral transform. We have performed the evolution by using Fourier Series, moments, and wavelets. We have also solved the evolution equations by discretizing the distribution. All approaches result in consistent results. The details of these algorithms will be described in a longer companion paper.

\begin{figure}[b]
  \centering
    \includegraphics[width=0.49\textwidth]{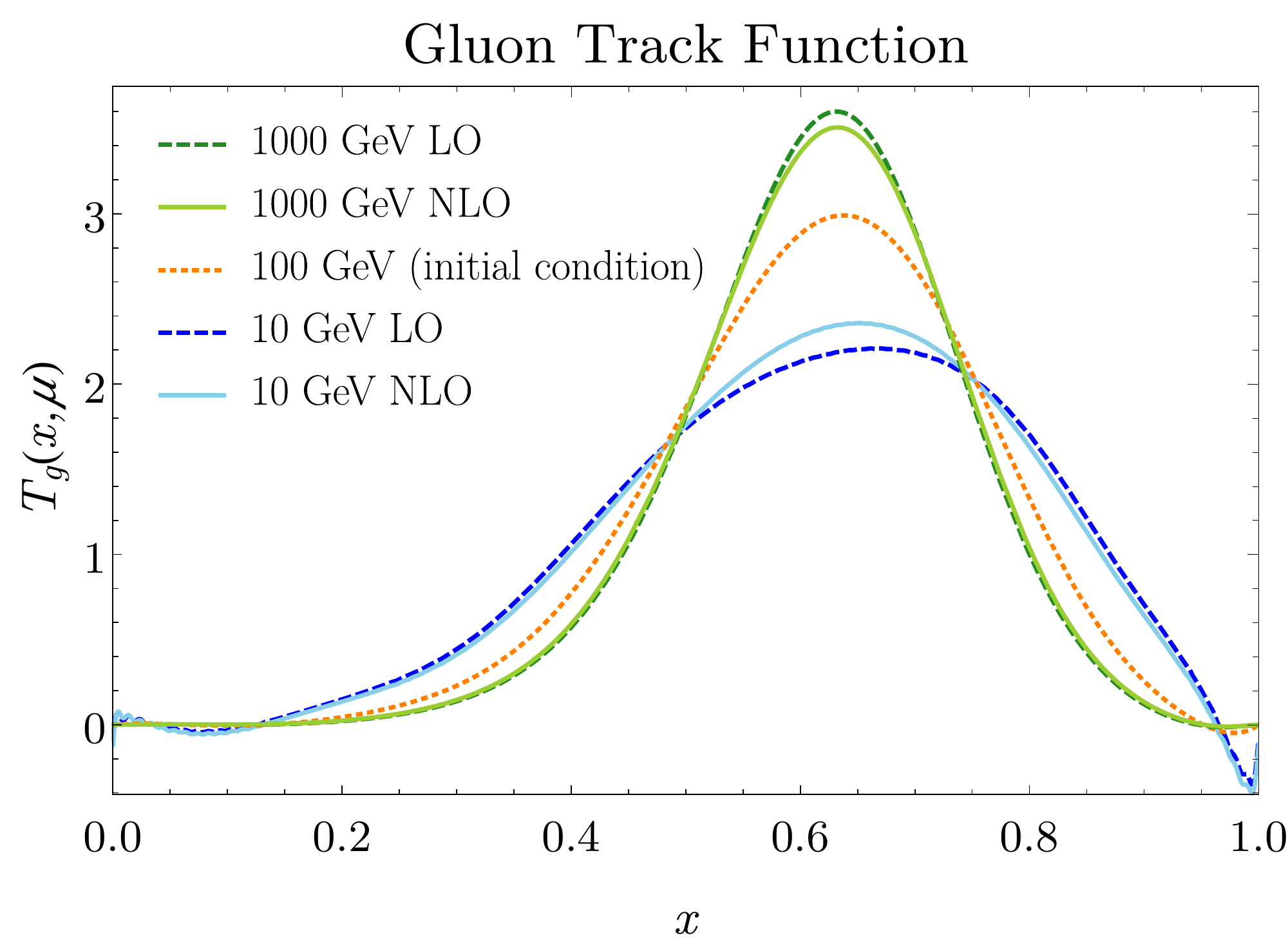}
  \caption{The evolution of the track function at LO and NLO. Initial conditions at 100 GeV are extracted from the Pythia parton shower. Higher order corrections are observed to be moderate in the peak of the distribution.}
  \label{fig:evolution}
\end{figure}

In \fig{evolution} we show the evolution of the track function distribution at both LO and NLO, considering the specific case of the track functions that measure the fraction of energy in charged particles. The initial condition  at $\mu=100$ GeV was extracted~\cite{Chang:2013rca} from Pythia~\cite{Sjostrand:2014zea}. The corrections from the NLO evolution are moderate, illustrating convergence. Our numerical algorithms, which are available at \url{https://github.com/HaoChern14/Track-Evolution}, allow the efficient evolution of track functions for phenomenological applications.

\emph{Applications.}---As an application of the evolution equation, we consider the calculation of the simplest physical observable depending on the track function, namely the fraction $w$ of energy in tracks in $e^+e^- \to$ hadrons. At lowest order, this object is essentially the convolution of track functions, $\int\! \df x\, T_q(x) T_{\bar q}(w-x)$, but it receives higher-order corrections. This can be viewed as an analog of the Drell-Yan process for illustrating the convergence of the DGLAP evolution for parton distribution functions. Furthermore, this observable, either at $e^+e^-$ colliders, or within jets at the LHC, can be used to extract the track functions from experimental data.

\begin{figure}
  \centering
     \includegraphics[width=0.49\textwidth]{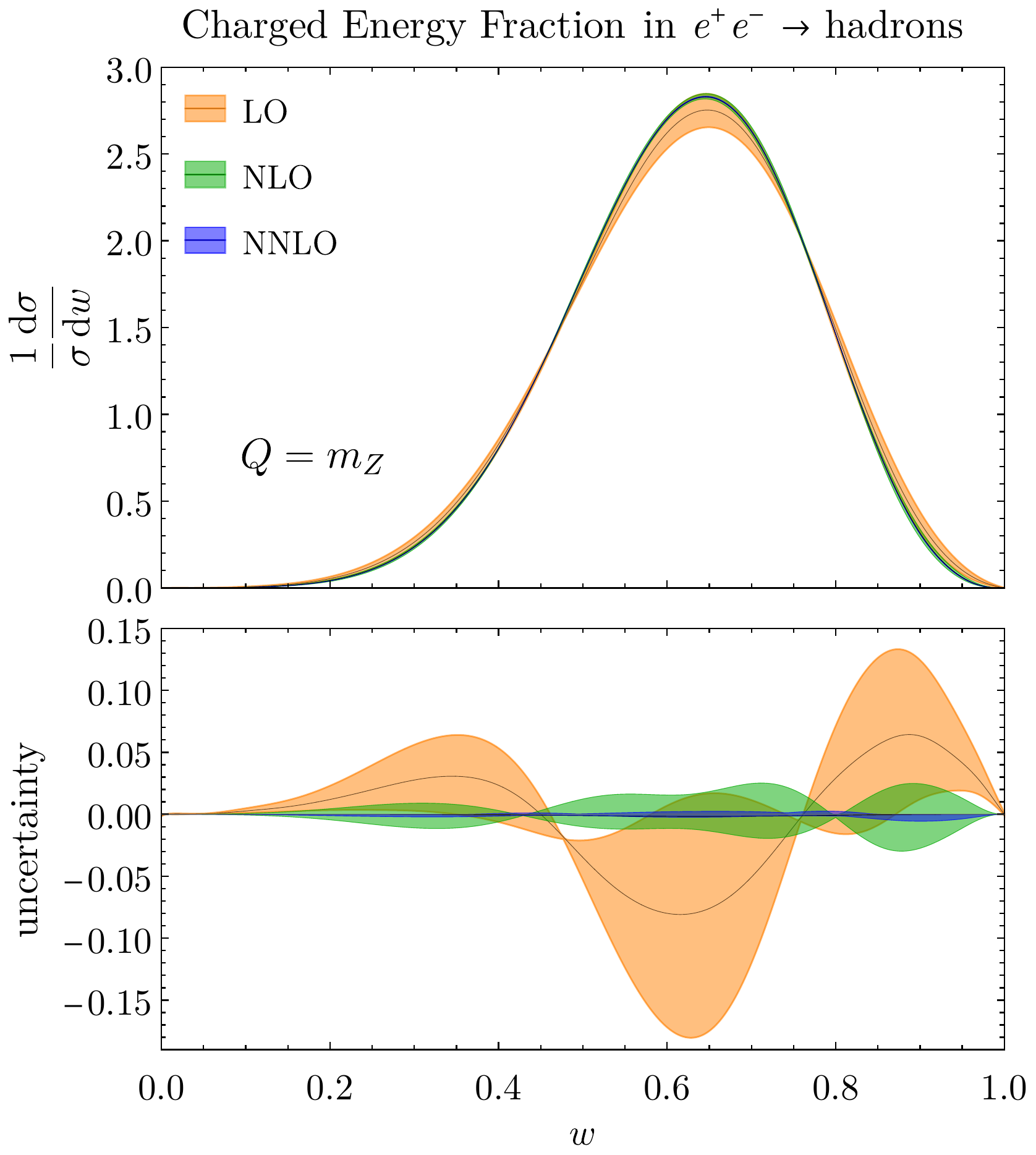} 
  \caption{A plot of the distribution of the energy fraction $w$ of charged particles produced in $e^+e^-$. In the bottom panel we show the difference to NLO to show the convergence of the perturbative series. A significant reduction in the uncertainty from scale variation is observed at NNLO due to including the NLO evolution of the track function. }
  \label{fig:wx}
\end{figure}

The distribution $w$ was first computed in \cite{Chang:2013rca} using the leading order evolution of the track functions. Here we are able to extend this to incorporate the NLO track function evolution, greatly reducing uncertainties from scale variation. In \fig{wx} we show the distribution of $w$ at LO, NLO and NNLO, along with the difference to the NLO result in the subpanel. At NNLO, we have only included logarithmic terms, not constants, so that the central value remains the same. We observe a large reduction in the scale uncertainty with the inclusion of the NLO evolution. We emphasize that this distribution requires the knowledge of the evolution of the complete track function, $T(x)$, not just its moments. This can also be applied to the measurement of the energy fraction of charged particles at the LHC, which is of interest for jet substructure.

\emph{Conclusions.}---In this \emph{Letter} we have gone beyond the DGLAP paradigm, by deriving non-linear collinear evolution equations incorporating multiparton correlations. We derived the explicit form of the evolution equations at NLO, which involve the $1\to3$ splitting functions, and showed how these equations can be solved numerically. As an application of these evolution equations we computed the distribution of energy in charged particles in $e^+e^-\to$ hadrons. 

The framework developed in this paper can be extended to higher orders, and all the amplitudes required to extend our approach to NNLO are available in the literature \cite{Badger:2004uk,Catani:2003vu,Bern:2004cz,Badger:2015cxa,Czakon:2022fqi,DelDuca:2019ggv,DelDuca:2020vst}. This would extend the description of the substructure of jets to the cutting edge of perturbative QCD.

With the central role that the DGLAP equations have played in understanding the dynamics of jets, we believe that our extension to include correlations in the fragmentation process will have a significant impact on many areas of jet physics, from the study of multi-hadron correlations, to improving the description of the perturbative substructure of jets, to the development of new parton shower algorithms. We look forward to their phenomenological application at the LHC and future colliders.

{\it Acknowledgements.}---
We thank Duff Neill for his knowledge of the literature, and Jingjing Pan for helpful discussions. 
H.C., Y.L.~and H.X.Z.~are supported by the Natural Science Foundation of China under contract No. 11975200.
M.J.~is supported by the NWO projectruimte 680-91-122. 
I.M. is supported by start-up funds from Yale University.
W.W is supported by the D-ITP consortium, a program of NWO that is funded by the Dutch Ministry of Education, Culture and Science (OCW). 

\bibliography{spinning_gluon.bib}{}

\setcounter{equation}{0}
\renewcommand{\theequation}{S-\arabic{equation}}

\end{document}